\documentclass[aps,twocolumn,epsf,showpacs,floatfix]{revtex4}
\usepackage{amssymb}
\usepackage{subfigure}
\usepackage{graphicx}
\usepackage[usenames]{color}

\begin{document}

\title{Symbiotic gap and semi-gap solitons in Bose-Einstein condensates}
\author{Sadhan K. Adhikari$^{1}$ and Boris A. Malomed$^{2}$}
\affiliation{$^1$Instituto de F\'{\i}sica Te\'{o}rica, UNESP -- S\~{a}o Paulo State
University, 01.405-900 S\~{a}o Paulo, S\~{a}o Paulo, Brazil \\
$^2$Department of Physical Electronics, School of Electrical Engineering,
Tel Aviv University, Tel Aviv 69978, Israel}

\begin{abstract}
Using the variational approximation (VA) and numerical simulations, we study
one-dimensional gap solitons in a binary Bose-Einstein condensate trapped in
an optical-lattice potential. We consider the case of inter-species
repulsion, while the intra-species interaction may be either repulsive or
attractive. Several types of gap solitons are found: symmetric or
asymmetric; 
unsplit or split, if centers of the components coincide or separate;
intra-gap (with both chemical potentials falling into a single bandgap) or
inter-gap, otherwise. 
In the case of the intra-species attraction, a smooth transition takes place
between solitons in the semi-infinite gap, the ones in the first finite
bandgap, and \textit{semi-gap} solitons (with one component in a bandgap and
the other in the semi-infinite gap).
\end{abstract}

\maketitle

\section{Introduction}

One of the milestones in studies of Bose-Einstein condensates (BECs) was the
creation of bright solitons in $^{7}$Li and $^{85}$Rb in ``cigar-shaped"
traps \cite{BECsolitons}, with the atomic scattering length made negative
(which corresponds to the attraction between atoms) by means of the
Feshbach-resonance (FR) technique \cite{solitons}. Normally, BEC features
repulsion among atoms. In that case, it was predicted that an
optical-lattice (OL) potential may support \textit{gap solitons} (GSs) \cite%
{GSprediction}, whose chemical potential falls in finite bandgaps of the
OL-induced spectrum. Although GSs, unlike ordinary solitons in
self-attractive BEC, cannot realize the ground state of the condensate, it
was demonstrated that they may easily be stable against small perturbations
\cite{stable}. A GS in $^{87}$Rb was experimentally created in a
cigar-shaped trap combined with an OL potential, pushing the BEC into the
appropriate bandgap by acceleration \cite{Markus}. Other possibilities for
the creation of GSs are offered by phase imprinting \cite{Veronica}, or
squeezing the system into a small region by a tight longitudinal parabolic
trap, which is subsequently relaxed \cite{Michal}.

BEC mixture of two hyperfine states of the same atom are also available to
the experiment \cite{binary}. The sign and strength of the inter-species
interaction may also be controlled by means of the FR \cite{inter-Feshbach},
hence one may consider a binary condensate with intra-species repulsion
combined with attraction between the species. It was proposed to use this
setting for the creation of \textit{symbiotic} solitons \cite{symbiotic}, in
which the attraction overcomes the intrinsic repulsion.

In this work, we aim to study compact (\textit{tightly bound }\cite{Arik})
\textit{symbiotic gap solitons} in a binary BEC, which are trapped,
essentially, in a single cell of the underlying OL potential. Unlike the
situation dealt with in Refs. \cite{symbiotic}, we consider the case of
inter-species repulsion, while the intra-species interactions may be
repulsive or attractive.
In Ref. \cite{Arik} it was already demonstrated that the addition of
intra-species repulsion expands the stability region of symbiotic GSs
supported primarily by the inter-species repulsion. The case of attraction
between two self-repulsive species was recently considered in Ref. \cite%
{Warsaw}, where it was shown that the attraction leads to a
counter-intuitive result -- \emph{splitting} between GSs formed in each
species. This effect can be explained by a\ negative effective mass, which
is a characteristic feature of the GS \cite{GSprediction}.
Indeed, considering the interaction of two GSs belonging to different
species, one may expect that the interplay of the attractive interaction
with the negative mass will split the GS pair.

Using variational \cite{VA} and numerical methods, we here construct
families of stable GSs of two kinds: unsplit (fully overlapping) and split
(separated). The splitting border is predicted by the variational
approximation (VA) in an almost exact form. In terms of chemical potentials
of the two components, the solitons may be of intra- and inter-gap types
\cite{Arik}, with the two components sitting, respectively, in the same gap
or different gaps. In particular, the states with one component residing in
the semi-infinite gap (which is possible in the case of intra-species
attraction) will be called semi-gap solitons.

The paper is organized as follows. The formulation of the system and
analytical results, obtained by the variational method \cite{VA}, are given
in Sec. II. Numerical findings are reported in Sec. III, including maps of
GS families in appropriate parameter planes. Section IV summarizes the work.


\section{Analytical considerations}

We consider a binary BEC loaded into a cigar-shaped trap combined with an OL
potential acting in the axial direction. Starting with the system of coupled
3D Gross-Pitaevskii equations (GPEs) for wave functions of the two
components, $\phi _{1}$ and $\phi _{2}$, one can reduce them to 1D equations
\cite{Luca}. In the scaled form, they are \cite{Warsaw}
\begin{eqnarray}
i\left( \phi _{1,2}\right) _{t} &=&-(1/2)\left( \phi _{1,2}\right) _{xx}+{\ g%
}|\phi _{1,2}|^{2}\phi _{1,2}  \nonumber \\
&+&{\ g_{12}}|\phi _{2,1}|^{2}\phi _{1,2}-V_{0}\cos \left( 2x\right) \phi
_{1,2},  \label{q1}
\end{eqnarray}%
where the OL period is fixed to be $\pi $, and the wave functions are
normalized to numbers of atoms in the two species, $\int_{-\infty }^{+\infty
}\left\vert \phi _{1,2}(x)\right\vert ^{2}dx=N_{1,2}$. In Eq. (\ref{q1}),
time, the OL strength, and nonlinearity coefficients are related to their
counterparts measured in physical units as follows: $t\equiv \left( \pi
/L\right) ^{2}\left( \hbar /m\right) t_{\mathrm{phys}}~,V_{0}\equiv \left(
L/\pi \hbar \right) ^{2}m(V_{0})_{{\mathrm{phys}}}$, $\left\{
g,g_{12}\right\} \equiv \left( 2Lm\omega _{\perp }/\pi \hbar \right) \left\{
a,a_{12}\right\} $, where $m$ is the atomic mass, $L$ the OL period, $a$ and
$a_{12}$ scattering lengths accounting for collisions between atoms
belonging to the same or different species, and $\omega _{\perp }$ the
transverse-confinement frequency. As said above, we assume repulsive
inter-species interactions, with $g_{12}>0$, while the intra-species
nonlinearity may be both repulsive ($g>0$) and attractive ($g<0$).

While the model assumes equal intra-species scattering lengths, they are, in
general, different for two hyperfine states \cite{binary}. Therefore, using
a FR, one cannot modify both intra-species nonlinearities to keep exactly
equal values of coefficient $g$ in equations for both components [cf. Eq. (%
\ref{q1})], running from negative to positive values (hence, strictly
speaking, different cases considered in this work cannot be realized in a
single mixture, but should be rather considered as a collection of
situations occurring in different mixtures). However, we will consider
asymmetric configurations, with $N_{1}\neq N_{2}$, which give rise to a much
stronger difference in the effective interaction strengths in the two
components than a small difference in their intrinsic scattering lengths.

Stationary solutions to Eqs. (\ref{q1}) are looked for in the usual form, $%
\phi _{1,2}(x,t)=\exp \left( -i\mu _{1,2}t\right) u_{1,2}(x)$, with chemical
potentials $\mu _{1,2}$ and functions $u_{1,2}(x)$ obeying
\begin{equation}
\mu _{1,2}u_{1,2}+u_{1,2}^{\prime \prime }/2-{\ g}u_{1,2}^{3}-{\ g_{12}}%
u_{2,1}^{2}u_{1,2}+V_{0}\cos \left( 2x\right) u_{1,2}=0,  \label{stationaryx}
\end{equation}%
with $\int_{-\infty }^{+\infty }u_{1,2}^{2}(x)dx=N_{1,2}$. In the GS
solutions constructed below, $\mu _{1}$ and $\mu _{2}$ belong to the first
two finite bandgaps and/or the semi-infinite gap in the spectrum induced by
potential $-V_{0}\cos \left( 2x\right) $.

\textit{Variational approximation for unsplit solitons:} Equation (\ref%
{stationaryx}) can be derived from Lagrangian
\begin{eqnarray}
L &=&\int_{-\infty }^{+\infty }\left[ \mu _{1}u_{1}^{2}+\mu _{2}u_{2}^{2}-%
\frac{1}{2}\left( \left( u_{1}^{^{\prime }}\right) ^{2}+\frac{1}{2}\left(
u_{2}^{^{\prime }}\right) ^{2}\right) \right.  \nonumber \\
&&+V_{0}\cos (2x)(u_{1}^{2}+u_{2}^{2})-\frac{1}{2}g\left(
u_{1}^{4}+u_{2}^{4}\right)  \nonumber \\
&&\left. -g_{12}u_{1}^{2}u_{2}^{2}\right] dx-\mu _{1}N_{1}-\mu _{2}N_{2}.
\label{L}
\end{eqnarray}%
To predict solitons with a compact symmetric profile, which corresponds to
numerical results displayed below, we adopt the Gaussian ansatz \cite{VA},
\begin{equation}
u_{1,2}^{\mathrm{(unsplit)}}(x)=\pi ^{-1/4}\sqrt{\frac{N_{1,2}\aleph _{1,2}}{%
w_{1,2}}}\exp \left( -\frac{x^{2}}{2w_{1,2}^{2}}\right) ,  \label{ansatz}
\end{equation}%
where variational parameters are widths $w_{1,2}$, reduced norms $\aleph
_{1,2}$, and $\mu _{1,2}$. The substitution of the ansatz in Eq. (\ref{L})
yields 
an effective Lagrangian, $L=L\left( \aleph _{1,2},w_{1,2},\mu _{1,2}\right) $%
. Then, the first pair of the variational equations, $\partial L/\partial
\mu _{1,2}=0$, gives $\aleph _{1,2}=1$, which is substituted below, after
performing the variation with respect to $\aleph _{1,2}$. Thus, the
remaining equations, $\partial L/\partial w_{1,2}=\partial L/\partial \aleph
_{1,2}=0$, take the form
\begin{equation}
1+\frac{gN_{1,2}w_{1,2}}{\sqrt{2\pi }}+\frac{2g_{12}N_{2,1}w_{1,2}^{4}}{%
\sqrt{\pi }(w_{1}^{2}+w_{2}^{2})^{3/2}}=4V_{0}w_{1,2}^{4}e^{-w_{1,2}^{2}}~,
\label{W1}
\end{equation}%
\begin{equation}
\mu _{1,2}=\frac{1}{4w_{1,2}^{2}}+\frac{gN_{1,2}}{\sqrt{2\pi }w_{1,2}}+\frac{%
g_{12}N_{2,1}}{\sqrt{\pi (w_{1}^{2}+w_{2}^{2})}}-V_{0}e^{-w_{1,2}^{2}}~.
\label{mu1}
\end{equation}%
Using Eqs. (\ref{W1}) and (\ref{mu1}) we can predict borders between
intra-gap and inter-gap soliton families of different types. To this end, we
take $\mu _{1,2}$ from Eqs. (\ref{mu1}) and, referring to the spectrum of
the linearized equation (\ref{q1}), identify curves in plane $\left(
N_{1},N_{2}\right) $ which correspond to boundaries between different gaps
in the two components.

\textit{Variational approximation for split solitons:} Two-component
solitons different from those considered above feature splitting between the
two components. An issue of obvious interest is to predict the splitting
threshold by means of the VA, for the symmetric case, with $%
N_{1}=N_{2}\equiv N$. For this purpose, we use the following ansatz,%
\begin{eqnarray}
&&u_{1,2}^{\mathrm{(split)}}(x)=\pi ^{-1/4}\sqrt{\frac{N}{w}}\left[ 1\pm bx+%
\frac{C}{4}w^{2}b^{2}\right.  \nonumber \\
&&\left. -\frac{1}{2}\left( 1+C\right) b^{2}x^{2}\right] \exp \left( -\frac{%
x^{2}}{2w^{2}}\right) ,  \label{A}
\end{eqnarray}%
with infinitesimal splitting parameter $b$, the objective being to find a
point at which a solution with $b\neq 0$ emerges. At small $b$, the two
components of expression (\ref{A}) feature maxima shifted to $x=\pm b/a+%
\mathcal{O}(b^{2})$, and up to order $b^{2}$, it satisfies the normalization
conditions, $\int_{-\infty }^{\infty }u_{1,2}^{2}(x)dx=N$. Unlike $b$,
constant $C$, to be defined below, is not a variational parameter.

The substitution of ansatz (\ref{A}) in Lagrangian (\ref{L}) yields, at
orders $b^{0}$ and $b^{2}$,
\begin{eqnarray}
&&L=-\frac{N}{2w^{2}}+2V_{0}e^{-w^{2}}-\frac{g+g_{12}}{\sqrt{2\pi }w}%
N^{2}-b^{2}N\left[ 1+\frac{C}{2}\right.  \nonumber \\
&&\left. -2CV_{0}w^{4}e^{-w^{2}}+\frac{g\left( C+2\right) +\left( C-2\right)
g_{12}}{2\sqrt{2\pi }}wN\right] .  \label{Leff}
\end{eqnarray}%
At order $b^{0}$ (i.e., for the unsplit soliton), variational equation $%
\partial L/\partial w=0$ reduces to Eq. (\ref{W1}) with $N_{1}=N_{2}\equiv N$
and $w_{1}=w_{2}\equiv w$:
\begin{equation}
1+\frac{\left( g+g_{12}\right) Nw}{\sqrt{2\pi }}=4V_{0}w^{4}e^{-w^{2}}.
\label{a}
\end{equation}%
At order $b^{2}$, equation $\partial L/\partial \left( b^{2}\right) =0$
yields the \textit{splitting condition},%
\begin{equation}
\frac{C+2}{4}\left( 1+\frac{gNw}{\sqrt{2\pi }}\right) +\frac{C-2}{4}\frac{%
g_{12}Nw}{\sqrt{2\pi }}-CV_{0}w^{4}e^{-w^{2}}=0.  \label{b}
\end{equation}%
Obviously, the splitting should not occur if $g_{12}=0$, i.e., Eq. (\ref{b})
must only yield the trivial solution, $w=0$, in this case. This condition
selects the value of $C$ which was arbitrary hitherto: $C=-2$, hence Eq. (%
\ref{b}) takes the form $Ng_{12}=2\sqrt{2\pi }V_{0}w^{3}e^{-w^{2}}$.
Combining this with Eq. (\ref{a}), we obtain $w=\sqrt{2\pi }/\left[ N\left(
g_{12}-g\right) \right] $, and a prediction for $N$ at the splitting point:%
\begin{equation}
N_{\mathrm{split}}^{4}=\frac{8\pi ^{2}V_{0}}{g_{12}\left( g_{12}-g\right)
^{3}}\exp \left[ -\frac{2\pi }{\left( g_{12}-g\right) ^{2}}\right] .
\label{g-gamma}
\end{equation}

\section{Numerical results}

\begin{figure}[tbp]
\begin{center}
{\includegraphics[width=.7\linewidth]{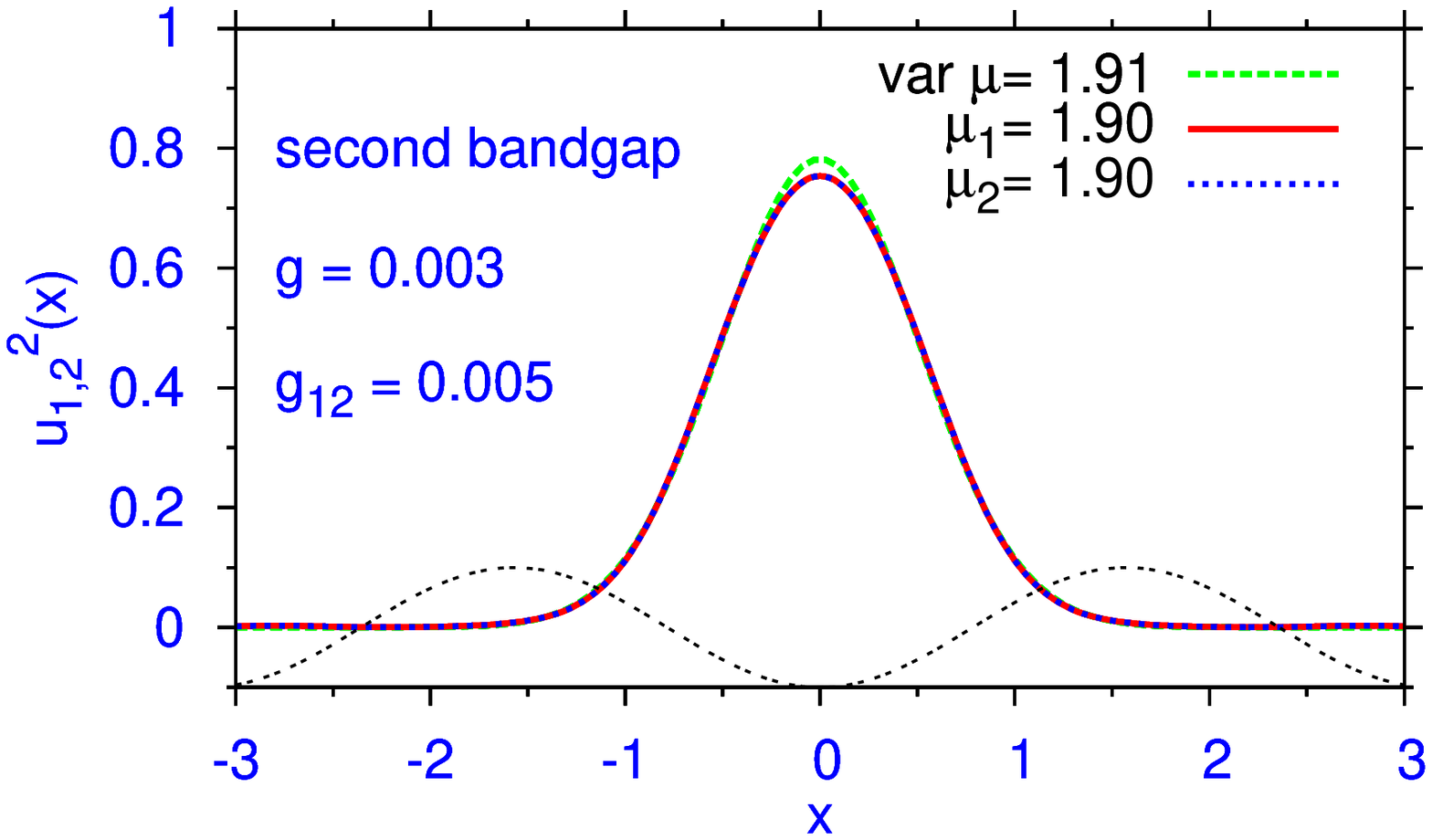}} {\includegraphics[width=.7%
\linewidth]{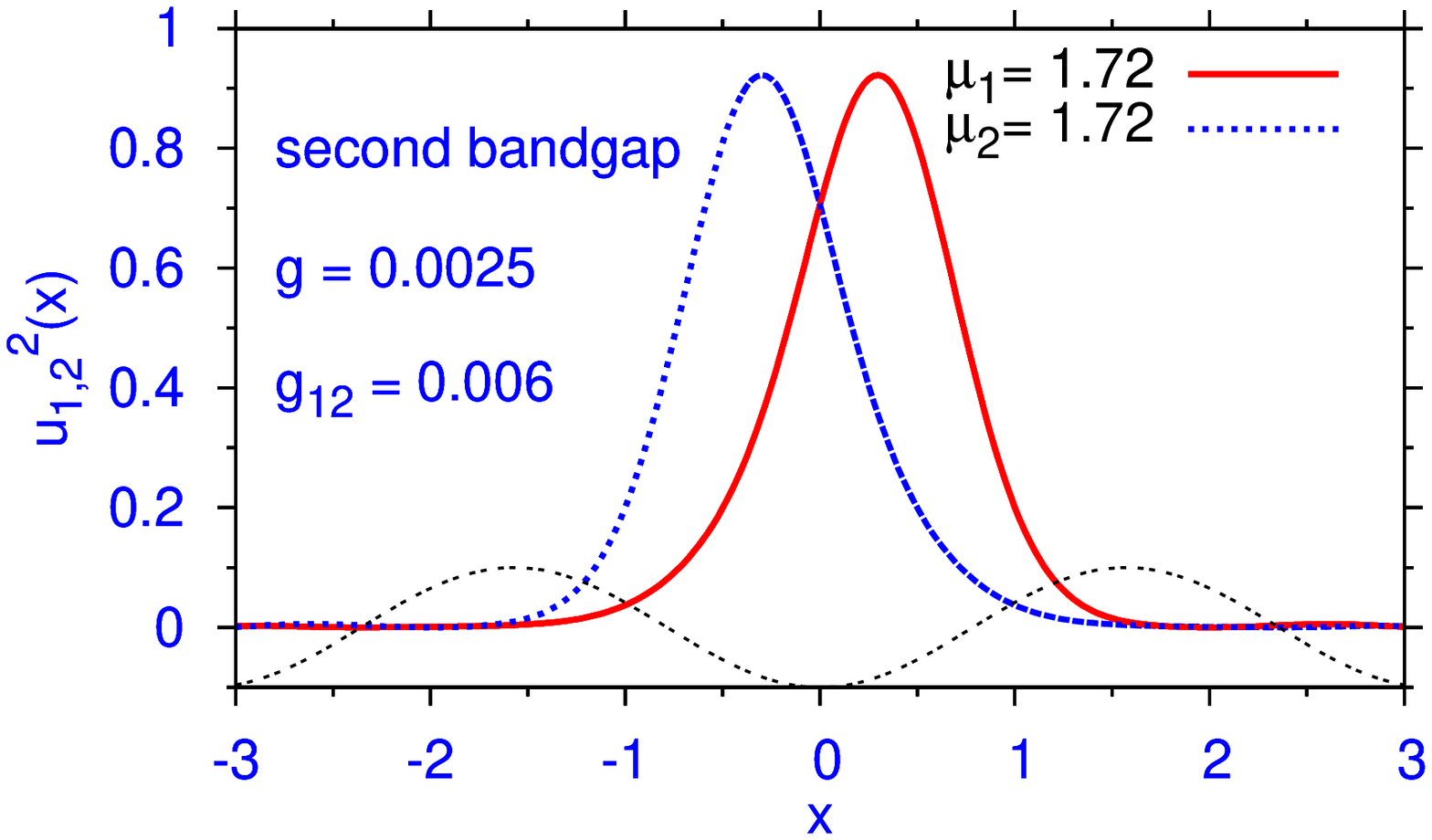}}
\end{center}
\caption{(Color online) Examples of unsplit and split symmetric solitons,
with $N_{1}=N_{2}=1000$, trapped in potential $-V_{0}\cos (2x)$. Here and in
all other figures, $V_{0}=5$. For the unsplit soliton, the variational
profile is included too.}
\label{examples-symm}
\end{figure}

\textit{Symmetric solitons:} Equation (\ref{q1}) was discretized using the
Crank-Nicholson scheme and solved numerically in real time, until the
solution would converge to a stationary soliton. This way of generating the
solitons guarantees their stability.
In Fig. \ref{examples-symm}, we present typical profiles of split and
unsplit symmetric solitons, with $N_{1}=N_{2}$. Due to the symmetry, these
solitons are always of the intra-gap type (in Fig. \ref{examples-symm}, they
belong to the second bandgap; in the semi-infinite and first gaps, the shape
of the solitons are quite similar).

In all cases, the difference between the variational and numerical shapes of
the unsplit solitons is extremely small. The present solitons are
essentially confined to a single cell of the OL potential. They change the
shape and develop undulating tails, which are often considered as a
characteristic feature of GSs, when $\mu $ is taken very close to an edge of
the bandgap (Fig. \ref{examples-symm} demonstrates that, even in a
well-pronounced split state, peaks of both components stay in a common
cell). It is also observed that, as might be expected, the increase of the
intra-species nonlinearity coefficient, $g$, pushes the solitons to higher
bandgaps, while the increase of $g_{12}$ tends to split the two components
of the soliton. In addition to the compact GSs presented here, there may
also exist loosely bound ones, that extend over several OL \cite{EPL}.

\begin{figure}[tbp]
{\includegraphics[width=.7\linewidth]{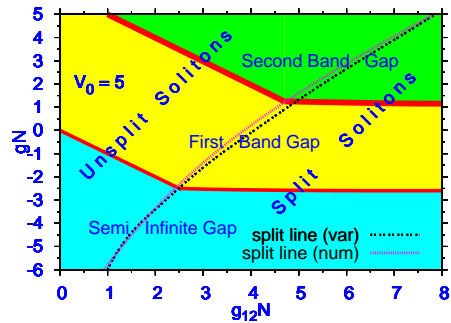}}
\caption{(Color online) The family of the two-component symmetric solitons ($%
N_{1}=N_{2}\equiv N$), mapped into the plane of the interaction strengths, $%
g_{12}N$ and $gN$. The plane is divided into regions corresponding to the
semi-infinite gap and the finite first and second bandgaps. They are
separated by narrow stripes representing the Bloch bands. The border between
the unsplit and split solitons is shown as found from the numerical data,
and as predicted by Eq. (\protect\ref{g-gamma}).}
\label{split-diagram}
\end{figure}

In Fig. \ref{split-diagram}, the entire family of the symmetric solitons is
displayed in the parameter plane of the strength of the intra- and
inter-species interactions, $\left( gN,g_{12}N\right) $. The VA prediction
for border between the unsplit and split solitons, given by Eq. (\ref%
{g-gamma}), provides a remarkably accurate fit to the numerical findings.

\begin{figure}[tbp]
\begin{center}
{\includegraphics[width=.7\linewidth]{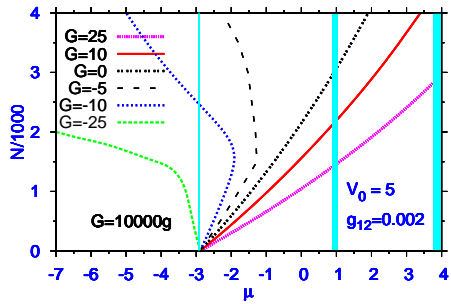}}
\end{center}
\caption{(Color online) The number of atoms in the symmetric soliton, $%
N_{1}=N_{2}\equiv N$, versus the common chemical potential of both
components, at several values of $g$ for $g_{12}=0.002$. Vertical stripes
are the Bloch bands between the gaps (the solution branch with $g=-0.001$
suffers a discontinuity when it hits the band separating the semi-infinite
and first finite gaps).}
\label{VKcriterion}
\end{figure}

Dependences $N(\mu )$ for families of symmetric solitons are plotted in Fig. %
\ref{VKcriterion}. It is known that a necessary stability condition for
solitons populating the semi-infinite gap is given by the Vakhitov-Kolokolov
(VK) criterion, $dN/d\mu <0$ \cite{VK}, while stable solitons in finite
bandgaps have $dN/d\mu >0$, disobeying this criterion \cite{GSprediction,EPL}%
. In the present case, Fig. \ref{VKcriterion} shows the same generic feature
(the semi-infinite gap contains solitons only for $g<0$, i.e., in the case
of the self-attraction). A noteworthy feature, \textit{viz}., a turning
point in dependence $N(\mu )$, is exhibited, for $g=-0.001$, by the solution
branch which passes from the semi-infinite gap into the first finite
bandgap, and also by the branch corresponding to $g=-0.0005$. Consequently,
two different \emph{stable} solitons can be found in the corresponding
interval of $\mu $. The solitons belonging to the branches with $g=0.0025$, $%
g=0.001$ and $g=0$ in Fig. \ref{VKcriterion} are unsplit, and they are
accurately predicted by the VA. Accordingly, the curves for these branches,
as obtained from the VA and from the numerical data, are virtually
identical. On the other hand, all solitons belonging to the branch with $%
g=-0.0025$ exhibit splitting. As concerns the bending branches, their parts
below the turning point are formed by unsplit solitons (which are accurately
approximated by the VA), while above the turning point the family continues
in the split form. Accordingly, the turning point on each bending branch
belongs to the splitting border for the symmetric solitons, cf. Fig. \ref%
{split-diagram}.

\begin{figure}[tbp]
\begin{center}
{\includegraphics[width=.7\linewidth]{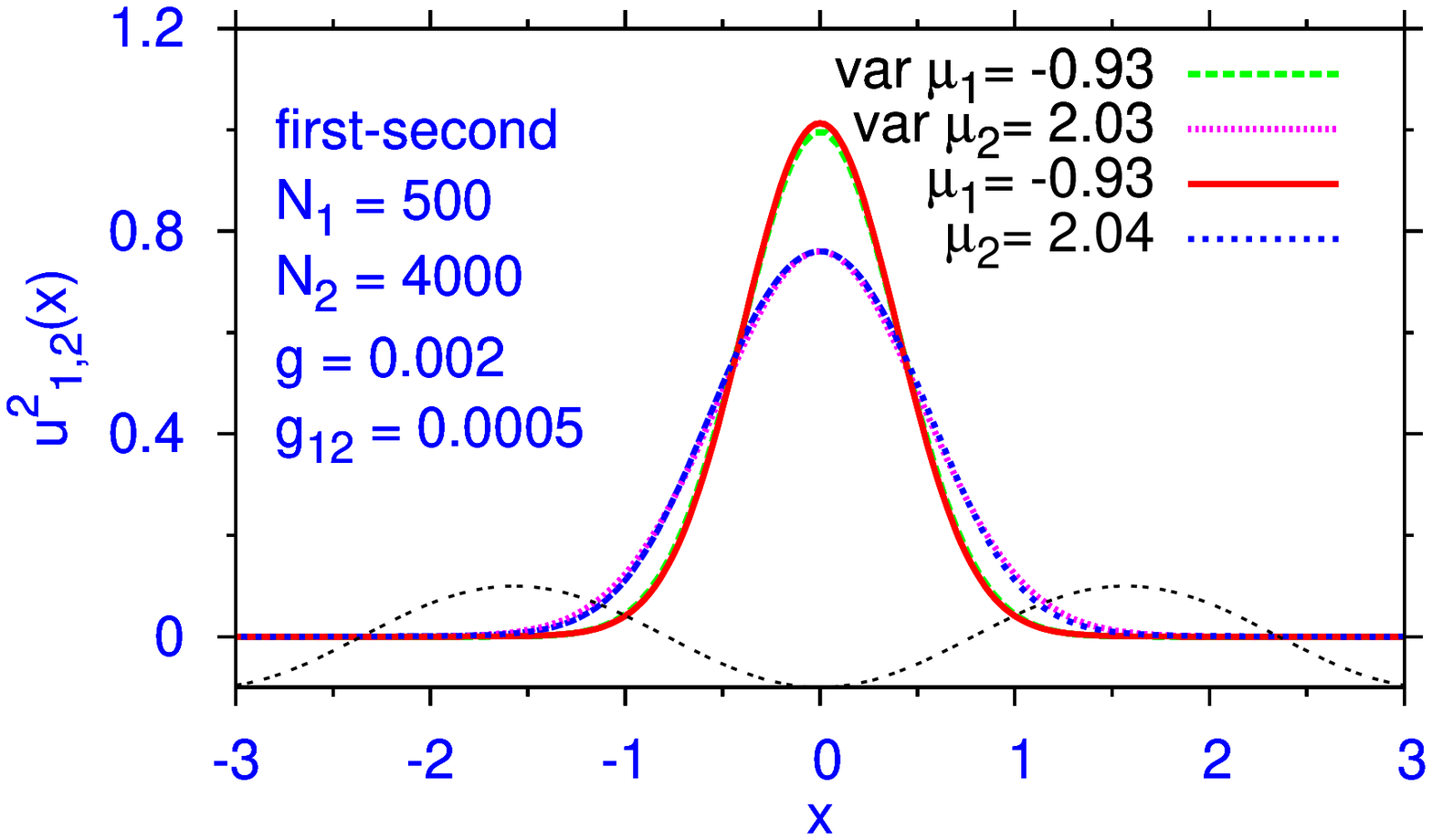}} {\includegraphics[width=.7%
\linewidth]{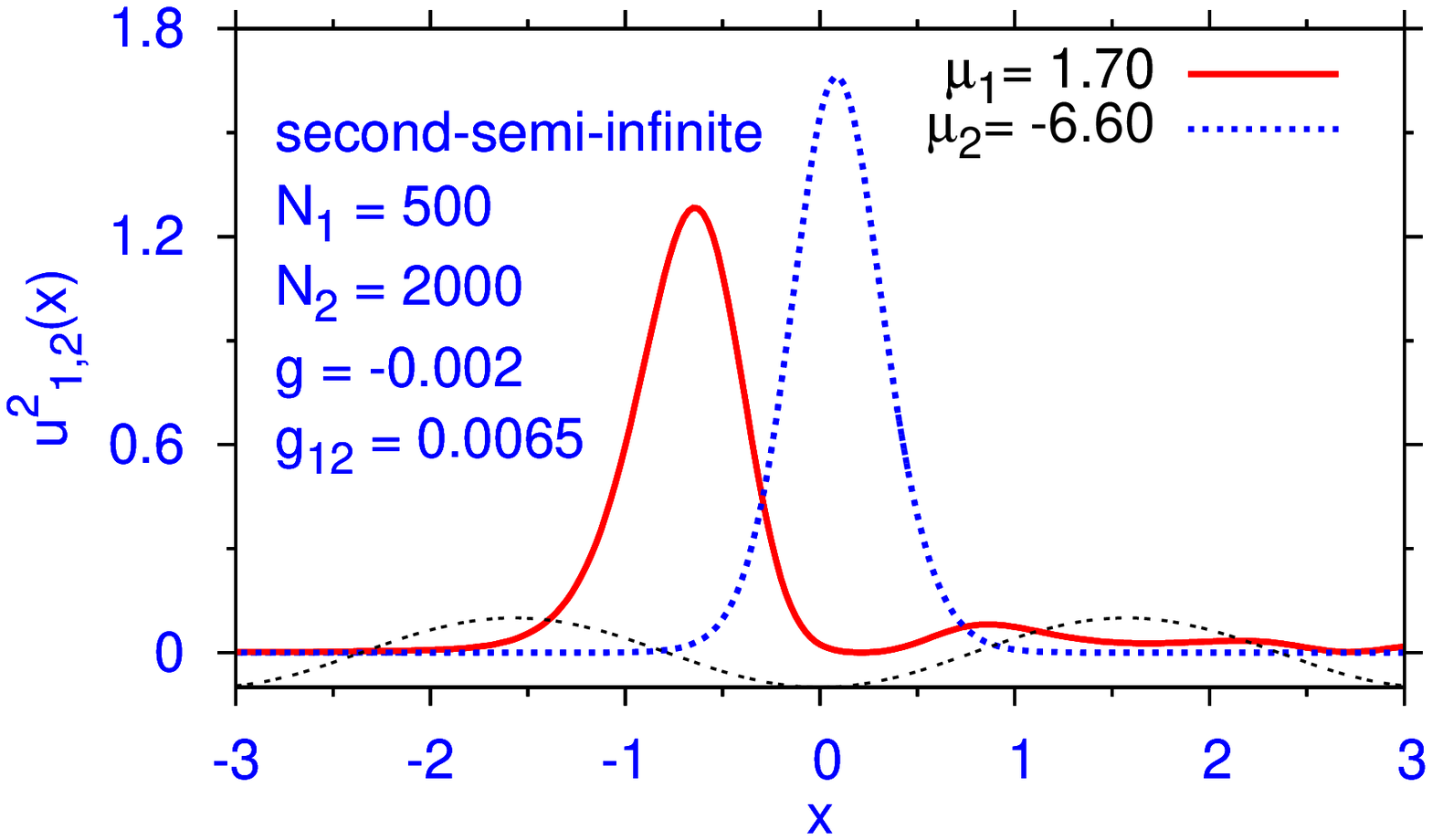}}
\end{center}
\caption{(Color online) Typical profiles of unsplit and split asymmetric ($%
N_{1}\neq N_{2}$) solitons. The examples represent solitons of inter-gap
types, as indicated in the panels. For the unsplit soliton, the profiles
predicted by the VA are shown too.}
\label{examples-asymm}
\end{figure}

\textit{Asymmetric solitons:} Typical examples of solitons with $N_{1}\neq
N_{2}$ are displayed in Fig. \ref{examples-asymm}. Similar to their
symmetric counterparts, cf. Fig. \ref{examples-symm}, they feature both
unsplit and split shapes (the former ones are well approximated by the VA),
which are again confined to a single cell of the OL potential.

\begin{figure}[tbp]
\begin{center}
{\includegraphics[width=.7\linewidth]{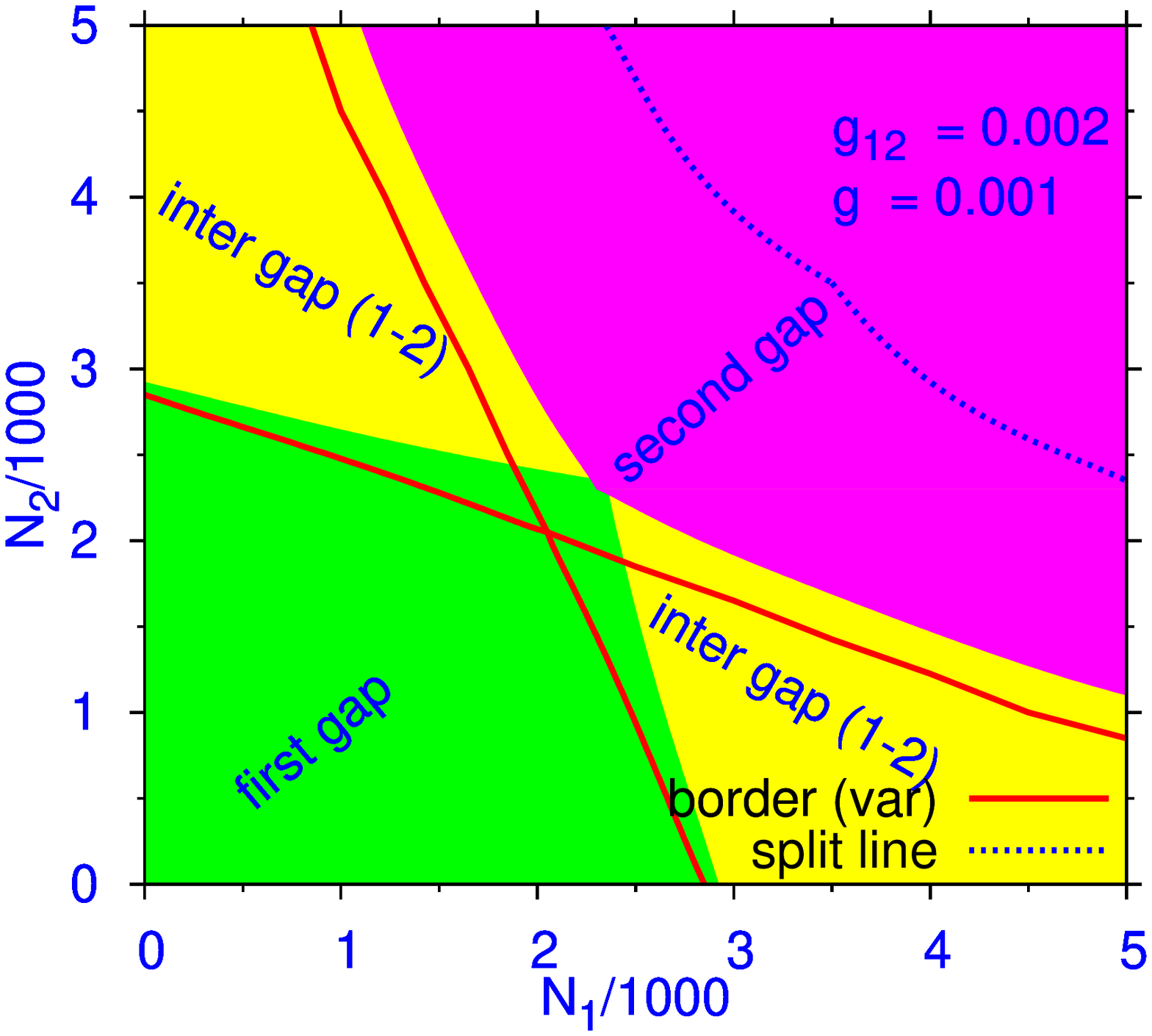}} {\includegraphics[width=.7%
\linewidth]{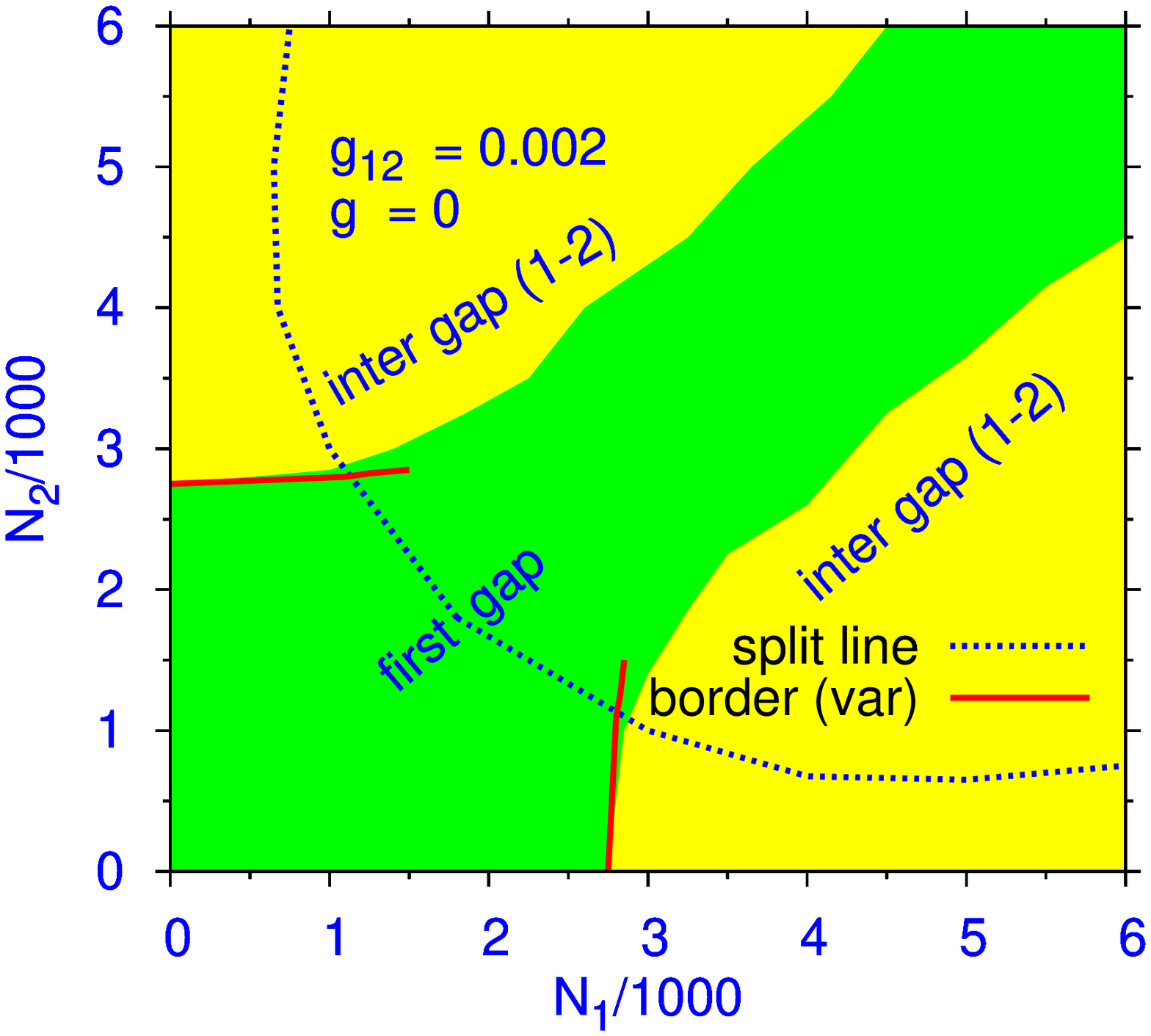}} {\includegraphics[width=.7\linewidth]{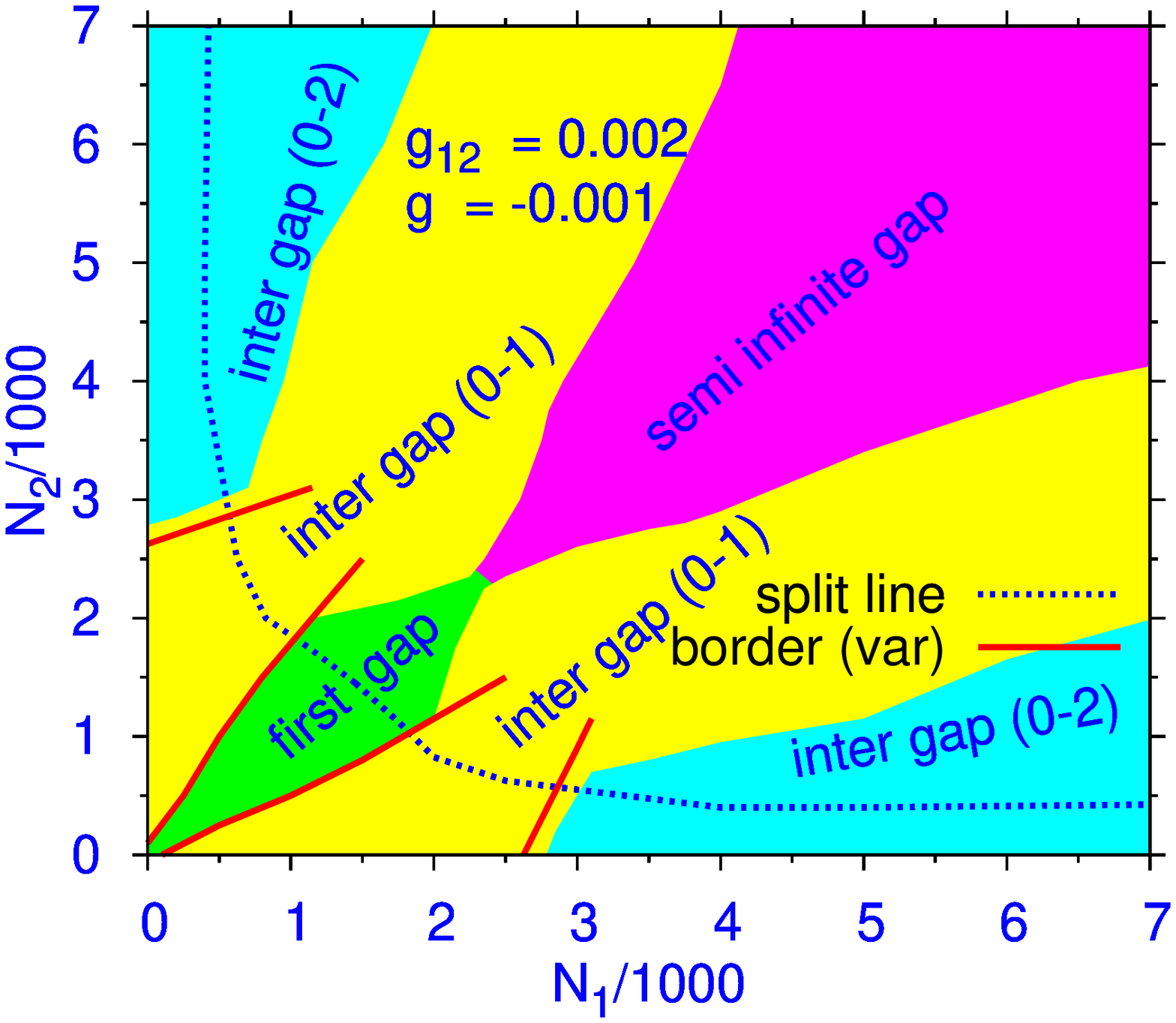}}
\end{center}
\caption{(Color online) Families of asymmetric and symmetric solitons mapped
into the plane of atom numbers $N_{1}$ and $N_{2}$, for different $g_{12}$
and $g$. The plane is divided into regions populated by solitons of six
different types (three intra-gap and three inter-gap varieties, symbols 0
and 1, 2 standing for the semi-infinite and two lowest finite bandgaps,
respectively). Each panel also shows the numerically found border between
the unsplit and split solitons, and borders between different types of the
unsplit ones, as predicted by the VA.}
\label{maps}
\end{figure}

The entire family of asymmetric and symmetric GSs is mapped in the $\left(
N_{1},N_{2}\right) $ plane, at fixed values of the interaction coefficients (%
$g_{12}$ and $g$), in Fig. \ref{maps}. In these diagrams, the border between
intra-gap solitons of different types shrink to a point belonging to the
diagonal line ($N_{1}=N_{2}$), which corresponds to symmetric solitons that
account for direct transitions between different types of intra-gap
solitons. In Fig. \ref{maps} the VA for the unsplit solitons accurately
predicts borders between their different varieties.

%

If none of the nonlinearities is attractive [Figs. \ref{maps}(a) and (b)],
no chemical potential may fall in the semi-infinite gap. Three types of GSs
are possible if both nonlinearities are repulsive [Fig. \ref{maps}(a)]:
intra-gap ones, in the two finite bandgaps, and the inter-gap species,
combining them. If the intra-species nonlinearity exactly vanishes [Figs. %
\ref{maps}(b)], the inter-species repulsion cannot push both components into
the second finite bandgap, which leaves us with two species: intra-gap in
the first bandgap, and the one mixing the two finite bandgaps. The interplay
of the attractive intra-species nonlinearity with the inter-species
repulsion supports two intra-gap and two inter-gap types, as seen in Fig. %
\ref{maps}(c). Note that one of them skips the first bandgap, binding
together components sitting in the semi-infinite and in second finite gaps.
A notable feature of the map in Fig. \ref{maps}(c) is the smooth transition
from ordinary solitons, with both components in the semi-infinite gap, to
ones of the semi-gap type.

\section{Conclusion}

In this work, we have considered the interplay of the repulsion between two
species of bosonic atoms with intra-species repulsion or attraction in a
binary BEC mixture loaded into the OL potential. Families of stable solitons
found in this setting are classified as symmetric/asymmetric, split/unsplit,
and intra/inter-gap. Three varieties of intra-gap solitons, and another
three types of inter-gap ones are identified, if the consideration is
limited to the two lowest finite bandgaps of the OL-induced spectrum.
Varying the atom numbers in the two components, $N_{1,2}$, we have plotted
maps of various states. Although different intra- and inter-gap species are
separated by Bloch bands, transitions between them are continuous in the $%
\left( N_{1},N_{2}\right) $ plane. In particular, a solution branch which
connects the solitons (of the split type), populating the semi-infinite gap,
and unsplit solitons in the first finite bandgap, features the turning point
at the border between the two varieties. Other varieties revealed by the
analysis represent semi-gap solitons, with one component belonging to the
semi-infinite gap, and the other one falling into a finite bandgap.

A considerable part of the numerical findings reported in this work was
accurately predicted by variational approximation. These include the shape
of unsplit solitons (both symmetric and asymmetric ones), borders between
their varieties, and the splitting border for the symmetric solitons.


We appreciate support from FAPESP and CNPq (Brazil), and Israel Science
Foundation (Center-of-Excellence grant No. 8006/03).

\end{document}